\documentclass[graybox]{svmult}

\usepackage{mathptmx}       % selects Times Roman as basic font
\usepackage{helvet}         % selects Helvetica as sans-serif font
\usepackage{courier}        % selects Courier as typewriter font
\usepackage{type1cm}        % activate if the above 3 fonts are
\usepackage{makeidx}         % allows index generation
\usepackage{graphicx}        % standard LaTeX graphics tool
\usepackage{multicol}        % used for the two-column index
\usepackage[bottom]{footmisc}% places footnotes at page bottom

\makeindex             % used for the subject index
\begin{document}

\title*{Time-resolved surveys of stellar clusters}
\author{Laurent Eyer, Patrick Eggenberger, Claudia Greco, Sophie Saesen, Richard I. Anderson, Nami Mowlavi}
\authorrunning{L. Eyer et al.}
\institute{Laurent Eyer, Patrick Eggenberger, Claudia Greco, Sophie Saesen, Richard I. Anderson \at Geneva Observatory, University of Geneva, CH-1290 Sauverny, \email{laurent.eyer@unige.ch}
\and Nami Mowlavi \at ISDC, Geneva Observatory, University of Geneva, CH-1290 Versoix}
\maketitle

\abstract*{We describe the information that can be gained when a survey is done multi-epoch, and its particular impact for open clusters. We first explain the irreplaceable information that multi-epoch observations are giving within astrometry, photometry and spectroscopy. Then we give three examples for results on open clusters from multi-epoch surveys, namely, the distance to the Pleiades, the angular momentum evolution of low mass stars and asteroseismology. Finally we mention several very large surveys, which are ongoing or planned for the future, Gaia, JASMINE, LSST, and VVV.}

\abstract{We describe the information that can be gained when a survey is done multi-epoch, and its particular impact for open clusters. We first explain the irreplaceable information that multi-epoch observations are giving within astrometry, photometry and spectroscopy. Then we give three examples for results on open clusters from multi-epoch surveys, namely, the distance to the Pleiades, the angular momentum evolution of low mass stars and asteroseismology. Finally we mention several very large surveys, which are ongoing or planned for the future, Gaia, JASMINE, LSST, and VVV.}

\section{Introduction}
\label{sec:Introduction}
The organizers of the conference asked us to present the following subject: What can time-resolved surveys of stellar clusters teach us? We show that time is an essential dimension to gain knowledge in astronomy when we want to learn about stars, clusters and the Galaxy. In this article, we focus on open clusters.

Astrophysics has some limited avenues to explore the Universe and its content. Some main trends for survey strategies are to observe:
\begin{itemize}
 \item deeper/fainter;
 \item wider areas of the sky (with as upper limit the whole sky);
 \item ``sharper", i.e., with a better resolution;
 \item in different wavelenghts;
 \item multi-epoch, i.e., to observe many times the same region of the sky.
\end{itemize}
In this text we explore the last item, noting that different astronomical subjects/objects may be explored/discovered depending on the above choices for a survey.

The three main pillars of astrophysics are astrometry, photometry and spectroscopy. We show in Table~\ref{Tab:single-multi-epoch} what can be gained by doing a multi-epoch survey with respect to single-epoch survey.

\begin{table}[bth]
\setlength{\tabcolsep}{3mm}
 \begin{center}
 \caption{Information from single- and multi-epoch surveys.} \vspace{5pt}
 \label{Tab:single-multi-epoch}
{
 \begin{tabular}{l l l l l}\hline
{\bf Meas. type} & {\bf Single-epoch} & {\bf Multi-epoch}             \\[1pt]
\hline
Astrometry      & position               &   parallax, proper motion     \\[2pt]
                & optical doubles        &   (projected) binary orbits   \\[2pt]
Photometry      & AstroParam (Teff, logg)&   variation of AstroParam     \\[2pt]
                & metallicity            &                               \\[2pt]
                & extinction$^*$         &                               \\[2pt]
                & age$^*$, distance$^*$  &                               \\[2pt]
                &                        &   variability types:           \\[2pt]
                &                        &   ecl. Bin/ planetary transits\\[2pt]                 
                &                        &   pulsation                   \\[2pt]
                &                        &   rotation                    \\[2pt]
                &                        &   eruptive phenomena, etc.    \\[2pt]
Spectroscopy    & AstroParam             &   variation of AstroParam     \\[2pt]
                & elemental abundance    &   line-profile variations     \\[2pt]
                & radial velocity        &   radial-velocity variations  \\[2pt]
                &                        &                               \\[2pt]
                & stellar rotation       &                               \\[2pt]
                \hline
* : clusters  &                        &                               \\[2pt]
 \end{tabular}
 }
\end{center}
\end{table}

We obtain the best astrophysical constraints when several observables are combined. For example bringing together the astrometric orbit and radial-velocity measurements, allows to solve entirely a binary system (inclination, masses, radii, semi-major axis, eccentricity); bringing together photometric and radial-velocity observations allows the determination of the radius of certain pulsating stars such as the Cepheids by the Baade-Wesselink method, etc.

It should be noted that even in single-epoch surveys, it is advisable to take multi-epoch data of some regions as it allows to establish the precision (internal errors) of the survey. For example, this has been done in SDDS and 2MASS.

\section{Use of multi-epoch observations for open clusters}

We have shown above that multi-epoch observations allow to derive many astrophysical quantities.
Some of these quantities require fairly elaborated work, like the modelling of eclipsing binaries. How do we know whether the models are correct?
If we consider that all stars of an open cluster have the same age and initial chemical composition, we can compare astrophysical quantities derived from measurements of different stars. Any significant difference between a star's property and the property of another star within the cluster or the cluster property is of interest and may point towards problems in stellar models or in the method employed to derive the quantity.
Open clusters are therefore unique tools to test the astrophysical models. 

Another interest of multi-epoch observations of open clusters is that a given property can be compared for clusters of different ages and metallicities. The idea is to see whether there is a dependence on metallicity or age, e.g., for the properties of variable star populations. Of particular interests are the instability strip boundaries, the fraction of variables within the instability strips, the variability amplitudes and periods, etc. (cf. Anderson et al., these proceedings).

We present in the following sections three applications of multi-epoch surveys for open clusters.

\subsection{Example 1: The distance to the Pleiades}
Thanks to its multi-epoch astrometric observations and derived parallaxes, Hipparcos data allowed to precisely determine and compare the distance of the Pleiades with the one obtained using the usual main sequence fitting technique. It resulted in a well-known mismatch: Hipparcos locates the Pleiades at $118.3 \pm 3.5$ pc \cite{vanLeeuwen1999}, and with the new reduction at $120\pm 3.5$ \cite{vanLeeuwen2009} closer than usually quoted results ($132 \pm 4$ pc, see e.g. \cite{Meynetetal1993}).
%The distance determination of Hipparcos is based on xx stars.

Various attempts have been undertaken to determine the distance, using the interferometric binary Atlas (\cite{Panetal2004}, \cite{Zwahlenetal2004}), the eclipsing binary HD~23642 discovered thanks to Hipparcos (cf.~\cite{VallsGabaud2007}, \cite{Munarietal2004}, \cite{Southworthetal2005}, \cite{Groenewegenetal2007}) and the Hubble Space Telescope to derive parallaxes of three star members of the Pleiades \cite{Soderblometal2005}. These studies give a distance between 132 to 139 pc, so larger than the Hipparcos result. We will not debate these results, but point out that all the methods were using multi-epoch observations, indeed most methods to derive distances are using the time domain.
Finally we note that the Gaia mission should resolve this debate.

\subsection{Example 2: Angular momentum evolution of low mass stars}
\label{subsection:HAT}
The angular momentum evolution of low mass stars can be studied thanks to multi-epoch photometric observations of open clusters.  One reason of the variability of low mass stars is the presence of spots on their surfaces. The period of the photometric variations thus directly yields the stellar rotation period. Observing open clusters of different ages (as derived from isochrone fitting) allows to describe the angular momentum evolution.  For a review of the subject we refer to \cite{IrwinBouvier2009} and Moraux and Bouvier (these proceedings). We therefore briefly present here the results of the HATNet project 
(Hungarian-made Automated Telescope network, cf.~\cite{Bakosetal2004}) dedicated to the
search of transiting exo-planets. These data allow other scientific
investigations, such as the period determination of F, G, K Pleiades stars \cite{Hartmanetal2010}, or general variability of K and M dwarf stars \cite{Hartmanetal2010b}. 
Thanks to HATNet data, 14 new Pleides members were discovered, and the number of known periods for the Pleiades stars has been increased by a factor of 5.  The HATNet results confirm previous indications that the spin-down stalls at $\simeq$100 Myr for the slowest rotating stars.  The HATNet results also reveal that inconsistencies remain for the radii, spectroscopic and photometric stellar spin rates for low mass stars. By this example we also want to show that large surveys with small telescopes (11-cm) can be scientifically very productive for open cluster science.

%\subsection{Example 2: Angular momentum evolution of low mass stars} 
%The angular momentum evolution of low mass stars can be studied thanks to multi-epoch photometric observations of open clusters. One reason of the variability of low mass stars is the presence of spots on their surface. The period of the photometric variations thus directly yields to the stellar rotation period. Observing open clusters of different ages allows to describe the angular momentum evolution. For a review of the subject we refer to \cite{IrwinBouvier2009} and Moraux and Bouvier (these proceedings). We therefore present briefly here the results of HAT (Hungarian-made Automated Telescope network, cf. \cite{Bakosetal2004}) dedicated to the search of exo-planets.  These data allow to perform other researches such as the period determination of F, G, K Pleiades stars \cite{Hartmanetal2010}. Thanks to HAT data, the number of known periods for these stars has been increased by a factor of 5. The HAT results confirm previous indications that the spin-down stalls at $\simeq$100 Myr for the slowest rotating stars. By this example we also want to show that large surveys with small telescopes (11-cm) can be scientifically very productive for open clusters science.

\subsection{Example 3: Observations of solar-like oscillations in stellar clusters}
\label{sec:kepler}
By directly obtaining observational constraints on the internal properties of stars, the study of stellar oscillation modes or asteroseismology is a valuable technique to improve our knowledge of the complex physical processes that take place in stellar interiors and to progress thereby in their modelling. The study of solar oscillations has provided a wealth of information on the internal structure of the Sun and stimulated various attempts to obtain similar observations for other stars. In past years, the spectrographs developed for exoplanet searches have achieved the accuracy needed to detect solar-like oscillations in other stars from the ground, while photometric measurements of solar-like oscillations are obtained from space thanks to the CoRoT (CNES/ESA) and the {\it Kepler} (NASA) missions. Solar-like oscillations are of course not restricted to solar-type stars but are expected in any star exhibiting a convective envelope able to excite acoustic waves. In particular, beautiful observations of solar-like oscillations in red giant stars have been recently obtained.

The wealth of information contained in these detections of solar-like oscillations for numerous red giants stimulated the theoretical study of the asteroseismic properties of red giants and also population studies aiming at reproducing the distribution of global asteroseismic properties for a large number of these stars. Red giants in clusters are particularly interesting targets since they enable to combine the capability of asteroseismic studies to probe stellar interiors with the valuable additional constraints resulting from the common origin of stars in clusters (same age and initial chemical composition). The first clear detection of solar-like oscillations for red giants in an open cluster has been recently reported \cite{Stello2010}. These observations were obtained for the open cluster NGC 6819 during the first 34 days of continuous science observations by {\it Kepler} using the spacecrafts long-cadence mode of about 30 minutes. These observations lead to the determination of the global asteroseismic properties and oscillation amplitudes for red-giant stars with different luminosities, which gives valuable constraints on the predicted scaling relations of these quantities with global stellar parameters. The asteroseismic measurements provide also additional tests for cluster membership. These preliminary results based on data sampled at the spacecrafts long cadence during about one month 
illustrate the valuable potential of solar-like observations in stellar clusters. Longer time series using the spacecrafts short cadence mode of about one minute will provide detection of oscillation modes in subgiant and turnoff stars. This should allow us to test important aspects of stellar evolution such as the mass-loss rate on the red giant branch \cite{Miglio2009b}. Since the rotational history of a star has a large impact on its global and asteroseismic properties during the red giant phase \cite{Eggenberger2010}, these observations will also provide valuable constrains on transport processes during the main sequence. Moreover,  NGC 6819 is not the only cluster that will be observed by {\it Kepler}, since there are four open clusters in {\it Kepler}'s field of view with different ages and metallicities. The observation of solar-like oscillations for stars in stellar clusters promise therefore great prospects for testing stellar evolution models and to progress in our knowledge of stellar physics.

\section{Some surveys}
There is an extremely large number of photometric surveys, past, present and planned. Here, we will do a somewhat unfair selection and consider only a few of them. First we discuss the Geneva open cluster survey aiming at the detection of stellar variability and we continue with four very large surveys with broader science cases. Open cluster science will probably benefit most when the data of these several projects will be combined.

\subsection{Geneva open cluster survey}
A long-term project devoted to the systematic search for variable stars in Galactic open clusters, was started at Geneva Observatory in 2002. Between 2002 and 2010, we observed 30 open clusters in both hemispheres using the 1.2-m Euler Swiss Telescope at La Silla and the 1.2-m Mercator Flemish telescope at La Palma, Spain. The open clusters in our sample range from metal-poor\footnote{Literature data taken from  WEBDA, \url{http://www.univie.ac.at/webda/}} (e.g. ${\langle {\rm [Fe/H]\rangle}} =-0.52 \ dex$, NGC~2324) to metal-rich (e.g. ${\langle {\rm [Fe/H]\rangle}} = 0.10 \ dex$, IC~4651), and from very young (e.g. 14 Myr, NGC~3766) to "old" (e.g. 2 Gyr, NGC~7789). Euler and Mercator share the same design, but have different fields of view: 11.5$^{\prime}$ $\times$ 11.5 $^{\prime}$ for Euler and 6.5$^{\prime}$ $\times$  6.5$^{\prime}$ for Mercator. The photometric observations were done in the Geneva $U$, $B$, and $V$ filters. The observations were scheduled so that the detection and phase coverage of both long and short period variables was optimized. All clusters were observed at least once per night during runs of two weeks, two to three times per year. We aimed to improve our coverage for short period variables by monitoring two clusters at higher cadence during each run. Observations of three of the southern clusters were limited to a 2-month baseline. In total, we obtained more than 2000 multi-band observations for each cluster.

\begin{figure}[th]
\sidecaption
\includegraphics[scale=.35]{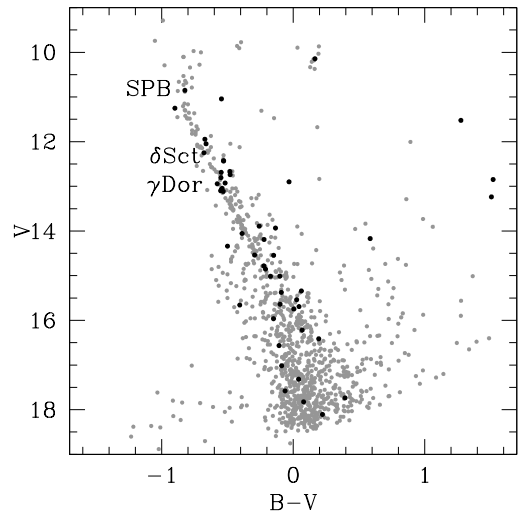}
\caption{Colour-Magnitude diagram of NGC~2447. We detect variability on time scales going from tens of minutes to tens of days in 54 variable stars in total. Below  14 $V$-mag, we find one slowly pulsating B star, one ellipsoidal binary, seven $\delta$ Sct stars, six $\gamma$ Dor stars and one hybrid $\delta$ Sct - $\gamma$ Dor pulsator.}
\label{fig:NGC2447}
\end{figure}

Preliminary results have been published for NGC~1901 \cite{Cherixetal2006}, NGC~5617 \cite{Carrieretal2009}, and additional publications are in preparation for IC~4651, NGC~2447, and NGC~2437. A preliminary colour magnitude diagram for NGC\,2447 including identified variables is shown in Fig.~\ref{fig:NGC2447}. We reach a precision of 5 mmag in $V$ for the brightest constant stars.
Reductions for the remaining clusters are ongoing. The methods used for the reduction of the data and the variability investigation are described in Greco et al. 2010 (in preparation) and \cite{Saesenetal2010}.

\subsection{Gaia}
Gaia is a space mission of the European Space Agency (for further information we refer to \url{http://www.rssd.esa.int/Gaia}). It will observe all objects brighter than V $\sim$ 20, recording the position, brightness, spectrophotometric and spectroscopic measurements, to determine distances, proper motions, stellar basic parameters and radial velocities. The mission duration is 5 years (with a possible extension of one year), during which the satellite will observe the entire sky an average of 70 times. One billion stars will be observed, with a tentative estimation of 100 million variables. The satellite will be launched from French Guyana in 2012. There will be intermediate data releases (though it is too early to develop the details of these releases) and the final results will be published in 2020-2021.

The scientific impact of Gaia will be tremendous in many fields of stellar physics, Galactic structure and Galactic history. Since it combines all the fundamental measurement types (astrometry, photometry, spectroscopy), it will also impact the subject of time-resolved science, particularly for open clusters.

Furthermore, if we want to unravel the formation history of our Galaxy, it has become clear that also detailed abundances play a crucial role. This is why across Europe there is an effort to get vast amounts of observing time from ESO spectrographs (which is included in the GREAT\footnote{http://www.ast.cam.ac.uk/GREAT/} initiative).

\subsection{JASMINE}
\label{subsection:JASMINE}
JASMINE is an acronym for Japan Astrometry Satellite Mission for INfrared Exploration. This project was planned in three phases of three satellites: Nano-JASMINE, Small-JASMINE, and JASMINE.

Nano-JASMINE (N-J), the first astrometric Japan satellite, should be launched in 2011 from Brazil by a Cyclone-4 rocket. N-J is fully funded by NAOJ (National Astronomical Observatory of Japan). The flight model will be completed by this December. The N-J catalogue will be opened after about 2 years operation. The observing strategy, whole sky scanning satellite, and method of the data analysis for N-J are similar to those for Gaia. N-J should reach in a zw-band (0.6‐1 $\mu$m) an astrometric precision of 2-3 mas. It will observe more than 10 million stars, complete to zw $\simeq$ 12.

Small-JASMINE (S-J) is a project that will observe in an infrared band (Hw-band:1.1--1.7micron). S-J will determine positions and parallaxes accurate to 10 $\mu$as and will have proper-motion errors of 9 $\mu$as/year for stars brighter than Hw=11.5 mag. It will observe small areas of the Galactic bulge with a single-beam telescope whose primary mirror diameter is around 30 cm. If selected by JAXA (Japan Aerospace Exploration Agency), the target launch date would be around 2016.

JASMINE is an extended mission of the S-J mission. It is designed to perform a survey towards the Galactic bulge region (20 by 10 degree$^2$) around the Galactic center with a single-beam telescope whose primary mirror diameter is around 80 cm. It will determine positions and parallaxes accurate to 10 $\mu$as and will have  proper-motion errors of 4~$\mu$as/year for stars brighter than Kw=11 mag. JASMINE would detect about one million bulge stars with the parallax uncertainty better than 10\%. If selected, JASMINE will be launched in the first half of the 2020s.

These projects should complement very well the Gaia mission, since Gaia will observe in the visible G-band and will be more sensitive to extinction. S-J and JASMINE will have J- and H-band photometry besides Hw-band, and Kw-band for astrometry, but the detailed design for photometry has not been determined yet.

\subsection{LSST}
LSST (cf. \cite{LSSTCol2009} and \url{http://www.lsst.org}) stands for Large Synoptic Survey Telescope. It is an ambitious project of an 8.4-m telescope that will be situated in Chile (Cerro Pachon).  The observations will consist of measurements of positions and of the 5 ugriz sloan$+$ y bands. During the 10 years of the project length, LSST will measure 1000 times half of the sky. The beginning of the observations is foreseen in 2017. The magnitude limit will attain 24.5 in a single image and 27 in stacked images. The number of objects that will be observed by LSST is estimated to be 10 billion stars and 10 billion galaxies. When performance and magnitude ranges are compared between Gaia and LSST, we remark there are many synergies.

\subsection{VVV}
We mention this ESO-survey because of its importance, but as it has been presented by Ivanov (details in these proceedings, see also \cite{Minnitietal2010}), we will therefore not develop it in this text. 

\begin{acknowledgement}
We would like to thank for their helpful comments Dr G.Bakos on section~\ref{subsection:HAT} and Prof. N.Gouda on section~\ref{subsection:JASMINE}.
\end{acknowledgement}
%\section{Conclusions}

\bibliographystyle{spphys}
\bibliography{eyer}

\begin{thebibliography}{10}
\providecommand{\url}[1]{{#1}}
\providecommand{\urlprefix}{URL }
\expandafter\ifx\csname urlstyle\endcsname\relax
  \providecommand{\doi}[1]{DOI \discretionary{}{}{}#1}\else
  \providecommand{\doi}{DOI \discretionary{}{}{}\begingroup
  \urlstyle{rm}\Url}\fi

\bibitem{vanLeeuwen1999}
F.~{van Leeuwen}, A\&A \textbf{341}, L71 (1999)

\bibitem{vanLeeuwen2009}
F.~{van Leeuwen}, A\&A \textbf{497}, 209 (2009)

\bibitem{Meynetetal1993}
G.~{Meynet}, et~al., A\&AS \textbf{98}, 477 (1993)

\bibitem{Panetal2004}
X.~{Pan}, et~al., Nature \textbf{427}, 326 (2004)

\bibitem{Zwahlenetal2004}
N.~{Zwahlen}, et~al., A\&A \textbf{425}, L45 (2004)

\bibitem{VallsGabaud2007}
D.~{Valls-Gabaud},  (2007), \emph{IAU Symposium}, vol. 240, pp. 281--289

\bibitem{Munarietal2004}
U.~{Munari}, et~al., A\&A \textbf{418}, L31 (2004)

\bibitem{Southworthetal2005}
J.~{Southworth}, et~al., A\&A \textbf{429}, 645 (2005)

\bibitem{Groenewegenetal2007}
M.A.T. {Groenewegen}, et~al., A\&A \textbf{463}, 579 (2007)

\bibitem{Soderblometal2005}
D.R. {Soderblom}, et~al., AJ \textbf{129}, 1616 (2005)

\bibitem{IrwinBouvier2009}
J.~{Irwin}, J.~{Bouvier},  (2009), \emph{IAU Symposium}, vol. 258, pp. 363--374

\bibitem{Bakosetal2004}
G.~{Bakos}, et~al., PASP \textbf{116}, 266 (2004)

\bibitem{Hartmanetal2010}
J.D. {Hartman}, et~al., MNRAS \textbf{408}, 475 (2010)

\bibitem{Hartmanetal2010b}
J.D. {Hartman}, G.{\'A}. {Bakos}, R.W. {Noyes}, B.~{Sip{\"o}cz},
  G.~{Kov{\'a}cs}, T.~{Mazeh}, A.~{Shporer}, A.~{P{\'a}l}, ArXiv e-prints
  (2009)

\bibitem{Stello2010}
D.~{Stello}, et~al., ApJL \textbf{713}, L182 (2010)

\bibitem{Miglio2009b}
A.~{Miglio}, et~al.,  (2009), \emph{AIPCS}, vol. 1170, pp. 132--136

\bibitem{Eggenberger2010}
P.~{Eggenberger}, et~al., A\&A \textbf{509}, A72+ (2010)

\bibitem{Cherixetal2006}
M.~{Cherix}, et~al., Memorie della Societa Astronomica Italiana \textbf{77},
  328 (2006)

\bibitem{Carrieretal2009}
F.~{Carrier}, et~al., Communications in Asteroseismology \textbf{158}, 199
  (2009)

\bibitem{Saesenetal2010}
S.~{Saesen}, et~al., A\&A \textbf{515}, A16+ (2010)

\bibitem{LSSTCol2009}
{LSST Science Collaborations}, et~al., ArXiv e-prints  (2009)

\bibitem{Minnitietal2010}
D.~{Minniti}, et~al., New Astronomy \textbf{15}, 433 (2010)

\end{thebibliography}
\end{document}